\documentclass[aps,prd,twocolumn,superscriptaddress,nofootinbib]{revtex4-2}

\usepackage[most]{tcolorbox}
\tcbuselibrary{breakable}
\usepackage[utf8]{inputenc}
\usepackage[T1]{fontenc}

\usepackage{amsmath,amssymb,amsfonts}
\usepackage{graphicx}
\usepackage{bm}
\usepackage{physics}
\usepackage{hyperref}
\usepackage{listings}
\usepackage{xcolor}

\usepackage{algpseudocode}

\usepackage{listings}
\usepackage{tcolorbox}

\usepackage{booktabs}
\usepackage{array}

\usepackage{tikz}
\usetikzlibrary{
	arrows.meta,
	positioning,
	calc,
	decorations.pathreplacing,
	shapes.geometric
}

\algrenewcommand\algorithmicrequire{\textbf{Input:}}
\algrenewcommand\algorithmicensure{\textbf{Output:}}

\algtext*{EndIf}
\algtext*{EndFor}
\algtext*{EndWhile}
\algtext*{EndProcedure}

\definecolor{codegreen}{rgb}{0,0.5,0}
\definecolor{codegray}{rgb}{0.5,0.5,0.5}
\definecolor{codepurple}{rgb}{0.58,0,0.82}
\definecolor{backcolour}{rgb}{0.96,0.96,0.96}

\lstdefinestyle{mystyle}{
	backgroundcolor=\color{backcolour},
	commentstyle=\color{codegreen},
	keywordstyle=\color{blue},
	numberstyle=\tiny\color{codegray},
	stringstyle=\color{codepurple},
	basicstyle=\ttfamily\scriptsize,
	breaklines=true,
	captionpos=b,
	keepspaces=true,
	numbers=left,
	numbersep=5pt,
	showspaces=false,
	showstringspaces=false,
	showtabs=false,
	tabsize=2
}

\begin{document}


\title{
	Quantum-Inspired Hamiltonian Optimization,
	Stochastic Tensor Networks
	and Adaptive Congestion Routing
	for Large-Scale QKD Networks
}

\author{José Luis Rosales}
\affiliation{
	Grupo de Investigación en Información y Computación Cuántica (GIICC),
	Universidad Politécnica de Madrid (UPM), Spain
}

\begin{abstract}
	
	Quantum Key Distribution (QKD) networks require routing methodologies
	capable of jointly optimizing latency, secret key generation rate,
	congestion, finite capacity and operational security constraints under
	dynamically evolving traffic conditions. In this work we introduce a quantum-inspired optimization framework for
	adaptive multi-demand routing in QKD communication networks based on
	effective Hamiltonian modelling, Quantum Monte Carlo inspired annealing
	and stochastic Tensor-Network State (TNS) compression.
	The communication network is represented as a stochastic interacting graph
	whose routing configurations evolve under an effective Hamiltonian
	containing latency, keyrate, congestion, risk and capacity terms.
	The resulting optimization landscape is explored through two complementary
	approaches: a stochastic Metropolis annealer based on incremental local
	Hamiltonian updates, and a stochastic boundary-MPS tensor-network
	approximation that compresses the low-energy routing sector through
	thermal branch selection.
	
	The resulting framework establishes a scalable bridge between
	QKD network orchestration, statistical-physics-inspired optimization,
	tensor-network compression and future quantum-native routing systems.
	
\end{abstract}
\maketitle
\section{Introduction}

Quantum Key Distribution (QKD) networks are expected to become one of the
fundamental technological layers of future quantum-secure communication
infrastructures \cite{
	Bennett84} -\cite{Martin25Holistic}. Unlike conventional packet-switched networks, QKD systems
must optimize not only classical routing quantities such as latency or path
availability, but also quantum-specific physical constraints, including
secret key generation rate, optical attenuation, finite channel capacity,
trusted-node exposure and operational security restrictions.

As QKD infrastructures evolve toward metropolitan, inter-city and
satellite-connected architectures, routing becomes a large-scale adaptive
optimization problem involving many simultaneous communication demands
competing for limited quantum resources. In this regime, the network no
longer behaves as a static shortest-path system. Congestion modifies the
future value of routing decisions, different demands interact indirectly
through shared communication links, and improving one quantity may degrade
another one. Minimizing latency, maximizing secret-key throughput and
reducing congestion therefore become coupled optimization objectives.

From a statistical-physics perspective, the routing problem naturally admits
an interpretation as an interacting Hamiltonian system. Each routing
configuration may be interpreted as a discrete many-body state whose quality
is encoded in an effective energy functional. Latency, risk and capacity
violations contribute positive energetic penalties, while high secret-key
generation rates contribute negative rewards. Congestion terms introduce
effective interactions between otherwise independent routing demands,
generating a rugged non-convex optimization landscape with metastable
configurations and frustration effects analogous to spin-glass systems
\cite{Mezard87,Binder86,Nishimori01}.

The present work develops this idea into a broad quantum-inspired framework
for QKD routing optimization. The first algorithmic component is a Quantum
Monte Carlo (QMC) inspired Metropolis annealer
based on stochastic thermal sampling and annealing dynamics
\cite{Metropolis53,Hastings70,Kirkpatrick83}. The method does not attempt
to simulate a microscopic quantum Hamiltonian; instead, it employs the logic
of statistical-mechanical thermal sampling and annealing to explore the
effective routing Hamiltonian. Local stochastic perturbations of the routing
state are proposed, sparse incremental Hamiltonian variations are computed,
and candidate configurations are accepted or rejected through a
temperature-dependent Metropolis rule.

An important conceptual aspect of the present work concerns the role
of quantum-inspired methodologies in routing optimization itself.
The algorithms developed here are implemented on classical computational
platforms and do not rely on fault-tolerant quantum hardware.
Nevertheless, the quantum-inspired formulation provides several important
advantages from both the modelling and algorithmic perspectives.

The primary benefit is the Hamiltonian reformulation of the routing
problem. Instead of treating routing as a collection of disconnected
heuristics, the communication network is interpreted as an interacting
energetic system in which latency, keyrate, congestion, risk and
capacity limitations become coupled contributions of a single effective
Hamiltonian. This representation naturally incorporates frustration,
competition and metastability effects generated by shared communication
resources, allowing the routing problem to be analysed using the language
of statistical physics and interacting many-body systems
\cite{Mezard87,Binder86,Nishimori01}.

Within this framework, Quantum Monte Carlo inspired annealing introduces
a controlled exploration mechanism capable of escaping locally optimal
routing configurations during early stages of the optimization process.
Similarly, tensor-network methods provide a compressed variational
description of the low-energy routing sector, allowing the effective
correlation structure of the routing landscape to be analysed through
the tensor bond dimension $\chi$
\cite{White92,White93,Schollwock11,Orus14,Verstraete08,Ferris12,Ferris14}.

An additional advantage of the quantum-inspired formulation is its
architectural portability. Once expressed as an effective Hamiltonian,
the same routing model becomes naturally compatible with Ising systems,
QUBO optimization, Quantum Annealing, QAOA-type variational circuits,
tensor-network optimization and hybrid quantum-classical workflows
\cite{Lucas14,Kadowaki98,Das08,Farhi14}. Even when executed on classical
hardware, the Hamiltonian abstraction therefore provides a unified
optimization language connecting quantum communication infrastructures,
statistical-physics-inspired optimization and future quantum-native
network orchestration systems.

At the same time, several reservations must be emphasized. The present
framework does not claim exponential quantum advantage, and the proposed
algorithms remain classical approximations of highly non-convex routing
landscapes. In strongly frustrated networks, stochastic annealing may
still become trapped in metastable configurations, while tensor-network
compression may require large bond dimensions when routing correlations
become too strong. The practical value of the quantum-inspired approach
therefore lies primarily in its modelling power, unified Hamiltonian
structure, adaptive optimization capability and compatibility with
future quantum-native computational paradigms.

The second component is a stochastic Tensor-Network State (TNS)
approximation inspired by Matrix Product States (MPS),
DMRG methods and tensor-network compression techniques
\cite{White92,White93,Schollwock11,Orus14,Verstraete08}. In this formulation, each communication demand behaves as a
discrete tensor degree of freedom selecting one route among a finite set of
candidate paths. Since the complete routing space scales exponentially with
the number of demands, the tensor-network algorithm compresses the
low-energy routing sector through boundary-MPS or tensor-train truncation.
Unlike purely deterministic tensor compression schemes, the present approach
introduces stochastic thermal branch survival and random demand ordering,
thereby producing a tensor-network dynamics directly comparable to the QMC
annealing process.

The coexistence of QMC and stochastic TNS formulations is particularly
useful because both methods probe different properties of the same routing
Hamiltonian. QMC explores the energy landscape dynamically through thermal
stochastic evolution, while TNS attempts to compress the low-energy sector
through variational tensor truncation. Their comparison therefore provides
information about the effective correlation structure of the routing
problem. If moderate tensor bond dimensions reproduce the same congestion
patterns obtained through QMC annealing, the routing landscape possesses a
compressible low-energy structure. Conversely, large tensor dimensions may
indicate strong congestion-induced correlations and high effective routing
complexity.

The framework is formulated independently of any particular programming
language or computational platform. The implementation principles are those
of modern scientific programming environments: sparse graph representations,
precomputed route observables, incremental load updates, sparse edge-index
routing structures and memory-aware local Hamiltonian calculations. This
allows the methodology to remain portable across symbolic, numerical and
high-performance computational ecosystems.

The article is organized as follows. Section~II introduces the QKD graph
model and the effective routing Hamiltonian. Section~III presents the
Quantum Monte Carlo inspired annealing dynamics. Section~IV discusses the
memory-aware implementation strategy and sparse incremental updates.
Section~V introduces adaptive congestion-aware rerouting. Section~VI
develops the QUBO and Ising reformulations together with their relation to
Quantum Annealing and QAOA architectures. Section~VII introduces the
stochastic tensor-network formulation and compares it with the QMC
approach. Finally, Section~VIII discusses future perspectives toward
quantum-native orchestration and intelligent autonomous routing systems.

Large-scale QKD infrastructures are expected to become
one of the foundational layers of future quantum internet
architectures \cite{Kimble08,Wehner18,Pirker18}.
\section{Quantum-Inspired Routing Framework}

\subsection{Graph-Based QKD Network Model}

The communication infrastructure is represented by an undirected graph
\begin{equation}
	G=(V,L),
\end{equation}
where $V$ is the set of network nodes and $L$ is the set of physical or logical
QKD communication links. If $n=|V|$ and $m=|L|$, the average connectivity is
\begin{equation}
	k=\frac{2m}{n}.
\end{equation}
Each link $e\in L$ carries a set of physical and operational quantities,
\begin{equation}
	\mathrm{link}(e)=\{\ell_e,\tau_e,r_e,c_e,\rho_e\},
\end{equation}
where $\ell_e$ is the optical distance, $\tau_e$ is the propagation latency,
$r_e$ is the secret key generation rate, $c_e$ is the effective link capacity
and $\rho_e$ is an operational or security risk factor.

The model assumes that latency is approximately proportional to optical
distance,
\begin{equation}
	\tau_e \propto \ell_e,
\end{equation}
whereas the secret key generation rate decreases approximately exponentially,
\begin{equation}
	r_e \sim e^{-\ell_e/\ell_0}.
\end{equation}
This captures the physical attenuation of quantum optical channels: photon
losses reduce the achievable secret key throughput and therefore modify the
effective routing value of long links.

\subsection{Multi-Demand Routing}

The network simultaneously supports $M$ communication demands. Each demand is
represented as
\begin{equation}
	d_a=(s_a,t_a,f_a),
\end{equation}
where $s_a$ is the source node, $t_a$ is the target node and $f_a$ is the
required flow. For each demand, $q$ candidate routing paths are generated. A
complete routing state is
\begin{equation}
	\mathbf p=(p_1,p_2,\ldots,p_M),
\end{equation}
where $p_a$ is an integer variable selecting one candidate path for demand
$a$. The total configuration space scales as
\begin{equation}
	N_{\mathrm{states}}=q^M,
\end{equation}
which rapidly becomes computationally prohibitive even for moderate values of
$M$ and $q$.

\subsection{Effective Routing Hamiltonian}

The central modelling step is to encode the routing problem as an effective
Hamiltonian,
\begin{equation}
	H=
	H_{\mathrm{lat}}
	+
	H_{\mathrm{key}}
	+
	H_{\mathrm{risk}}
	+
	H_{\mathrm{cong}}
	+
	H_{\mathrm{cap}}.
\end{equation}
The latency term penalizes slow routes,
\begin{equation}
	H_{\mathrm{lat}}=
	\sum_a
	\alpha_{\mathrm{lat}}\tau_a,
\end{equation}
whereas the keyrate term rewards routes with high secret-key throughput,
\begin{equation}
	H_{\mathrm{key}}=
	-\sum_a
	\beta_{\mathrm{key}} r_a.
\end{equation}
The negative sign transforms keyrate maximization into energy minimization.
Operational risk contributes as
\begin{equation}
	H_{\mathrm{risk}}=
	\sum_a
	\gamma_{\mathrm{risk}}\rho_a.
\end{equation}

The interaction between demands is generated by link sharing. If
$\chi_{a,e}=1$ when demand $a$ uses link $e$ and zero otherwise, the link load
is
\begin{equation}
	\mathrm{load}(e)=
	\sum_a f_a\chi_{a,e}.
\end{equation}
The congestion energy is then
\begin{equation}
	H_{\mathrm{cong}}=
	\lambda
	\sum_e \mathrm{load}(e)^2,
\end{equation}
and capacity overload is penalized by
\begin{equation}
	H_{\mathrm{cap}}=
	\mu
	\sum_e
	\max(0,\mathrm{load}(e)-c_e)^2.
\end{equation}
These two terms are responsible for the effective many-body structure of the
routing problem, because the cost of assigning one demand to a path depends on
the paths already selected by other demands.

\subsection{Model Parameters and Programming-Environment Variables}

The Hamiltonian model contains three conceptually different parameter levels.
The first level consists of network-generated quantities, including link
lengths $\ell_e$, latencies $\tau_e$, key generation rates $r_e$, capacities
$c_e$, risk factors $\rho_e$, demand flows $f_a$ and candidate paths. These
variables describe the simulated or measured QKD network. The second level
contains Hamiltonian weights such as $\alpha_{\mathrm{lat}}$,
$\beta_{\mathrm{rate}}$, $\gamma_{\mathrm{risk}}$,
$\lambda_{\mathrm{cong}}$ and $\mu_{\mathrm{cap}}$; these are optimization
hyperparameters that define the engineering preference among latency,
keyrate, risk and congestion. The third level contains algorithmic parameters,
including the annealing schedule, the number of Monte Carlo steps, the tensor
bond dimension $\chi$, the tensor temperature and the stochastic truncation
noise.

In a programming environment, the simulation begins by defining the number of
nodes $n$, the average connectivity $k$ and the approximate number of links,
\begin{equation}
	m\simeq \frac{nk}{2}.
\end{equation}
A connected random graph is generated so that feasible routes exist between
the selected source and target nodes. Each physical link receives a length
$\ell_e$ sampled from a prescribed interval, for example
$\ell_e\in[5,40]$ km in a simplified metropolitan QKD scenario. The latency is
computed as
\begin{equation}
	\tau_e=4.9\,\ell_e,
\end{equation}
where the numerical coefficient may be interpreted as a propagation delay in
microseconds per kilometre. The secret key generation rate can be modelled as
\begin{equation}
	r_e=100 e^{-\ell_e/25}+\eta_e,
\end{equation}
with $\eta_e$ a small random fluctuation. The capacity is generated as
\begin{equation}
	c_e=r_e+\xi_e,
	\qquad
	\xi_e\in[10,30],
\end{equation}
so that better physical links tend to support larger effective traffic, while
still retaining stochastic variability. The risk factor is sampled as
$\rho_e\in[0.05,0.4]$ and may encode reliability, exposure, trusted-node
sensitivity, maintenance status or administrative security classification.

For every demand $a$ and candidate route $p$, the simulation precomputes
route-level observables. The route latency is
\begin{equation}
	\tau_{a,p}
	=
	\sum_{e\in \mathrm{route}(a,p)}
	\tau_e,
\end{equation}
the route keyrate is limited by the weakest link,
\begin{equation}
	r_{a,p}
	=
	\min_{e\in \mathrm{route}(a,p)}
	r_e,
\end{equation}
the route capacity is
\begin{equation}
	c_{a,p}
	=
	\min_{e\in \mathrm{route}(a,p)}
	c_e,
\end{equation}
and the route risk is
\begin{equation}
	\rho_{a,p}
	=
	\sum_{e\in \mathrm{route}(a,p)}
	\rho_e.
\end{equation}
These quantities are stored as numerical arrays and remain fixed during the
optimization. The candidate paths themselves are stored as sparse integer
edge-index lists, which avoids manipulating symbolic graph objects during the
Monte Carlo or tensor-network dynamics.

Because latency, keyrate and risk have different physical units and numerical
scales, the route observables are normalized:
\begin{widetext}
\begin{equation}
	\tilde{\tau}_{a,p}
	=
	\frac{\tau_{a,p}}{\max_{a,p}\tau_{a,p}},
	\qquad
	\tilde{r}_{a,p}
	=
	\frac{r_{a,p}}{\max_{a,p}r_{a,p}},
	\qquad
	\tilde{\rho}_{a,p}
	=
	\frac{\rho_{a,p}}{\max_{a,p}\rho_{a,p}}.
\end{equation}
\end{widetext}
The local energy of selecting route $p$ for demand $a$ is then
\begin{align}
	h_{a,p}
	=&\;
	\alpha_{\mathrm{lat}}
	\tilde{\tau}_{a,p}
	-
	\beta_{\mathrm{rate}}
	\tilde{r}_{a,p}
	+
	\gamma_{\mathrm{risk}}
	\tilde{\rho}_{a,p}
	\nonumber\\
	&+
	\mu_{\mathrm{cap}}
	\left[
	\max(0,f_a-r_{a,p})^2
	+
	\max(0,f_a-c_{a,p})^2
	\right].
\end{align}
The complete energy of a routing state is therefore
\begin{equation}
	H(\mathbf p)
	=
	\sum_{a=1}^{M} h_{a,p_a}
	+
	\sum_{e\in L}
	\Phi_e(\mathrm{load}(e)),
\end{equation}
with
\begin{equation}
	\Phi_e(x)
	=
	\lambda_{\mathrm{cong}}x^2
	+
	\mu_{\mathrm{cap}}\max(0,x-c_e)^2.
\end{equation}

\section{Quantum Monte Carlo Inspired Optimization}

The effective routing Hamiltonian defines a non-convex energy landscape with
many local minima generated by competing demands and congestion interactions.
A purely greedy optimization method would rapidly become trapped in
metastable configurations. The Quantum Monte Carlo inspired strategy instead
uses stochastic thermal dynamics to explore the configuration space.

At each Monte Carlo iteration, the algorithm selects one demand $a$ and
proposes a local change in its route,
\begin{equation}
	p_a\rightarrow p'_a.
\end{equation}
The energy variation is
\begin{equation}
	\Delta H=
	H(\mathbf p')-H(\mathbf p),
\end{equation}
but this variation is computed incrementally. Only the local route energy and
the congestion penalties of the links removed from, or added to, the selected
path are updated. The move is accepted with Metropolis probability
\begin{equation}
	P_{\mathrm{accept}}
	=
	\min(1,e^{-\beta\Delta H}).
\end{equation}

At the beginning of the annealing process, the inverse temperature $\beta$ is
small and energetically unfavourable moves are accepted with significant
probability, allowing exploration of the landscape. At the end, $\beta$ is
large and the dynamics becomes selective. A typical geometric schedule is
\begin{equation}
	\beta(t)
	=
	\beta_0
	\left(
	\frac{\beta_F}{\beta_0}
	\right)^{
		\frac{t-1}{N_{\mathrm{steps}}-1}
	}.
\end{equation}
This process is analogous to thermal relaxation in spin-glass systems: the
routing state fluctuates strongly at high temperature and gradually
concentrates around low-energy configurations.

The method is quantum-inspired rather than a full microscopic quantum
simulation. Its quantum character lies in the Hamiltonian formulation, the
analogy with interacting spin systems, and the use of stochastic sampling
methods closely related to the statistical-mechanical intuition underlying
Quantum Monte Carlo. In the present context, this is sufficient to use QKD
routing as a physically meaningful testbed for annealed Hamiltonian
optimization.

\section{Memory-Aware Implementation in Programming Environments}

One of the practical contributions of the framework is that the routing
dynamics can be simulated efficiently in general scientific programming
environments. Although the full configuration space scales as $q^M$, the
Hamiltonian has a sparse local-update structure. Each Monte Carlo move affects
only one demand and a small number of links, so the global energy does not need
to be recomputed from scratch.

The implementation is based on three principles. First, candidate routes are
converted into integer edge-index lists. Second, route observables such as
latency, keyrate, capacity and risk are precomputed and stored as numerical
arrays. Third, the current link-load vector is treated as a persistent
structure updated incrementally after accepted moves. With this design, a
single move requires work proportional to the average route length $L$ rather
than to the total number of links $E$,
\begin{equation}
	O(E)\rightarrow O(L),
	\qquad L\ll E.
\end{equation}

\begin{tcolorbox}[colback=gray!10,colframe=black,
	title=Meta-Code 1: Incremental Monte Carlo Update]
\begin{lstlisting}[language=Python]
choose random demand a
choose alternative route p'
compute local route-energy difference
remove old-route load from affected links
add new-route load to affected links
compute local congestion-energy difference
accept or reject with Metropolis criterion
\end{lstlisting}
\end{tcolorbox}

\begin{tcolorbox}[colback=gray!10,colframe=black,
	title=Meta-Code 2: Sparse Route Representation]
\begin{lstlisting}[language=Python]
route[a][p] -> [e1, e2, e3, ...]
load[e]     -> current flow crossing link e
energy      -> local route terms + link penalties
\end{lstlisting}
\end{tcolorbox}

This design is independent of the particular language used to implement the
model. In symbolic notebook environments the emphasis is on avoiding large
symbolic objects during the optimization; in numerical environments the same
logic becomes sparse arrays, packed numerical vectors and compiled local
updates; in high-performance environments it maps naturally to low-level array
kernels or parallel route proposals.

\section{Adaptive Congestion-Aware Routing}

After the global Monte Carlo optimization, the framework introduces a
secondary adaptive routing layer. Its purpose is to route an additional QKD
message between nodes $A$ and $B$ by minimizing the marginal congestion energy
rather than the topological path length.

For a new flow $f_{\mathrm{new}}$, the marginal cost of using link $e$ is
\begin{equation}
	\Delta H_e
	=
	\Phi^{\mathrm{marg}}_e
	\left(\mathrm{load}(e)+f_{\mathrm{new}}\right)
	-
	\Phi^{\mathrm{marg}}_e
	\left(\mathrm{load}(e)\right)
	+
	\epsilon,
\end{equation}
where
\begin{equation}
	\Phi^{\rm marg}_e(x)
	=
	\lambda_{\rm marg}x^2
	+
	\mu_{\rm ov}
	\max(0,x-c_e)^2 .
\end{equation}
The small positive $\epsilon$ prevents exactly zero path weights and stabilizes
the shortest-path computation. The weighted graph defined by $\Delta H_e$ is
then used to compute the minimum-congestion route. This mechanism produces a
form of autonomous network adaptation analogous to intelligent SDN routing:
the best route for a new message depends on the instantaneous energetic state
of the QKD network.

\begin{tcolorbox}[colback=gray!10,colframe=black,
	title=Meta-Code 3: Adaptive Congestion-Aware Routing]
\begin{lstlisting}[language=Python]
read current optimized load vector
for each link compute marginal congestion cost
construct weighted congestion graph
compute minimum-congestion path from A to B
update network load if the new flow is accepted
\end{lstlisting}
\end{tcolorbox}

\section{Relation to Ising, QUBO and Quantum Annealing}

The same routing Hamiltonian can be reformulated as an Ising or QUBO system
commonly employed in combinatorial optimization and quantum annealing
\cite{Lucas14,Kadowaki98,Das08}.
Introduce binary variables
\begin{equation}
	x_{a,p}\in\{0,1\},
\end{equation}
where $x_{a,p}=1$ means that route $p$ is selected for demand $a$. The
one-route-per-demand constraint is
\begin{equation}
	\sum_{p=1}^{q}x_{a,p}=1,
\end{equation}
which can be enforced energetically through a quadratic penalty. The QUBO
Hamiltonian has the form
\begin{equation}
	H_{\mathrm{QUBO}}
	=
	\sum_i h_i x_i
	+
	\sum_{ij}J_{ij}x_i x_j.
\end{equation}
Here, $h_i$ are local route biases generated by latency, keyrate, risk and
local capacity penalties, while $J_{ij}$ are interaction couplings induced by
shared links and congestion. If two routes share network resources, their
simultaneous activation produces an energetic interaction.

Binary variables can be mapped into Ising spins through
\begin{equation}
	x_i=\frac{1+\sigma_i}{2},
	\qquad
	\sigma_i\in\{-1,+1\}.
\end{equation}
The effective Ising Hamiltonian is
\begin{equation}
	H_{\mathrm{Ising}}
	=
	H_0+
	\sum_i h_i\sigma_i+
	\sum_{ij}J_{ij}\sigma_i\sigma_j.
\end{equation}
This establishes a direct conceptual connection between QKD routing and sparse
interacting spin systems.

In a quantum annealing interpretation, the system evolves from an initial
transverse-field Hamiltonian
\begin{equation}
	H_{\mathrm{initial}}
	=
	-\Gamma \sum_i \sigma_i^x
\end{equation}
toward the problem Hamiltonian,
\begin{equation}
	H_{\mathrm{problem}}=H_{\mathrm{QUBO}},
\end{equation}
according to
\begin{equation}
	H(t)=A(t)H_{\mathrm{initial}}+B(t)H_{\mathrm{problem}},
\end{equation}
where $A(t)$ decreases and $B(t)$ increases. The same QUBO structure can also
be used as the cost Hamiltonian in QAOA-type variational circuits, with mixer
terms designed to preserve or penalize the one-route-per-demand constraint\cite{Farhi14}.

\section{Stochastic Tensor-Network Formulation}

The Quantum Monte Carlo strategy explores the routing configuration space
through stochastic thermal moves. Tensor-Network States provide a
complementary variational-compression approach
widely employed in quantum many-body physics
and correlated systems
\cite{Orus14,Schollwock11,Verstraete08}: rather than walking through the landscape, they try to
compress its low-energy sector. This is motivated by the exponential scaling
\begin{equation}
	N_{\mathrm{states}}=q^M.
\end{equation}
Each demand is treated as a discrete tensor index,
\begin{equation}
	p_a\in\{1,\ldots,q\},
\end{equation}
and a complete routing configuration is the many-body state
\begin{equation}
	\mathbf p=(p_1,p_2,\ldots,p_M).
\end{equation}

Recent years have also seen the emergence
of stochastic tensor-network techniques
and tensor Monte Carlo methods
for sampling and compressing large correlated systems
\cite{Ferris12,Ferris14}.
The formal matrix-product representation of an amplitude or score over
routing configurations can be written as
\begin{equation}
	\Psi(p_1,p_2,\ldots,p_M)
	\approx
	A^{[1]}_{p_1}
	A^{[2]}_{p_2}
	\cdots
	A^{[M]}_{p_M},
\end{equation}
where the internal tensor bonds encode correlations between routing demands.
The bond dimension $\chi$ controls how much correlation structure is retained.
In the routing problem, these correlations are induced physically by shared
links and congestion bottlenecks.

For practical optimization, the implementation uses a boundary-MPS or
tensor-train truncation algorithm. The demands are incorporated sequentially.
At an intermediate step, the algorithm stores a compressed set of partial
states,
\begin{equation}
	\mathcal S_k=
	\{s_1,s_2,\ldots,s_\chi\},
\end{equation}
where each partial state contains an assignment of already-processed demands,
a current load vector and an accumulated energy. When the next demand is
added, every retained branch is expanded over all $q$ possible routes. The
expanded set is then compressed back to at most $\chi$ branches.

A purely deterministic truncation keeps the lowest-energy branches at every
step. This is fast, but it can become too rigid: once the graph, demands and
candidate routes are fixed, the same branches tend to survive. The stochastic
TNS variant avoids this by assigning a Boltzmann-like survival weight to each
expanded branch,
\begin{equation}
	w_i=
	\exp\left[
	-\beta_{\mathrm{TNS}}(E_i-E_{\min})
	\right],
\end{equation}
and a normalized survival probability
\begin{equation}
	P_i=\frac{w_i}{\sum_j w_j}.
\end{equation}
Branches are sampled probabilistically according to $P_i$, optionally with a
small numerical noise
\begin{equation}
	E_i\rightarrow E_i+\eta_i,
	\qquad
	\eta_i\in[-\epsilon_{\mathrm{TNS}},\epsilon_{\mathrm{TNS}}],
\end{equation}
which breaks near-degeneracies without changing the intended physical
Hamiltonian.

The tensor inverse temperature can follow its own annealing schedule,
\begin{equation}
	\beta_{\mathrm{TNS}}(k)
	=
	\beta_{0,\mathrm{TNS}}
	\left(
	\frac{\beta_{F,\mathrm{TNS}}}{\beta_{0,\mathrm{TNS}}}
	\right)^{
		\frac{k-1}{M-1}
	}.
\end{equation}
Early tensor steps retain more diverse partial configurations; later steps
become more selective. A random demand ordering
\begin{equation}
	\pi=(\pi_1,\pi_2,\ldots,\pi_M)
\end{equation}
adds a further source of stochasticity and changes the effective tensor-chain
structure, which is important because tensor truncation is order dependent.

\begin{tcolorbox}[colback=gray!10,colframe=black,
	title=Meta-Code 4: Stochastic Boundary-MPS Routing Optimizer]
\begin{lstlisting}[language=Python]
initialize graph, demands and candidate routes
precompute local routing energies
sample random demand order
initialize empty tensor boundary

for tensor step k = 1,...,M:
    select demand a = demandOrder[k]
    expand every retained branch over q routes
    update link loads incrementally
    compute local Hamiltonian increment
    compute tensor inverse temperature betaTNS[k]
    assign Boltzmann survival weights
    sample candidate branches stochastically
    remove duplicates and retain chi branches

select the lowest-energy final branch
reconstruct state in the original demand order
compute final load and routing observables
\end{lstlisting}
\end{tcolorbox}

The bond dimension $\chi$ is the central accuracy parameter. If $\chi=q^M$,
the method becomes exact but intractable. For moderate $\chi$, the algorithm
acts as a compressed approximation to the low-energy routing sector. The
approximation is expected to work well when congestion interactions are sparse
and correlation lengths among routing decisions remain moderate; it becomes
harder when many demands compete for the same bottleneck links, producing high
effective correlation complexity.

\section{Comparison Between QMC and Stochastic TNS}

QMC and stochastic TNS minimize the same Hamiltonian, but they do so through
different algorithmic mechanisms. QMC maintains one complete routing state at
a time and evolves it through local stochastic moves accepted with the
Metropolis rule. TNS maintains a compressed set of partial states and evolves
a boundary through stochastic branch expansion and truncation. Thus, QMC
performs stochastic exploration in time, while TNS performs stochastic
compression in tensor space.

This distinction makes the comparison scientifically useful. Energy alone is
not sufficient, because two routing states may have similar Hamiltonian values
but different load distributions. For QKD networks, the spatial structure of
load, the location of congestion hotspots, the number of overloaded links and
the behaviour of adaptive rerouting are as important as the scalar energy.
Both algorithms should therefore be compared through final energy, selected
routes, load-per-link distribution, congested links, highlighted used links,
minimum-congestion rerouting between nodes $A$ and $B$, and the difference
between topological shortest path and congestion-aware path.

If QMC and TNS agree, the routing landscape has a stable low-energy structure
that is both thermally discoverable and tensor-compressible. If they disagree
but have comparable energies, the landscape contains degeneracy and
frustration. If TNS requires a large $\chi$ to approach QMC performance, the
network exhibits high effective correlation complexity, meaning that routing
decisions cannot be compressed easily into a low-bond-dimension tensor
boundary.

\section{Perspectives for Quantum Machine Learning}

The routing simulations naturally generate structured datasets: network
topologies, link observables, demand matrices, selected paths, Hamiltonian
histories, congestion distributions, metastable configurations and adaptive
rerouting decisions. These datasets can support future machine learning
architectures for autonomous QKD orchestration.

Graph Neural Networks could learn to predict congestion hotspots, bottleneck
links, route quality or failure-prone regions directly from graph features.
Reinforcement Learning could learn adaptive routing policies through
interaction with the evolving network state. From the quantum-computing side,
the Hamiltonian formulation can be embedded into variational quantum circuits,
QAOA layers, quantum kernels or hybrid quantum-classical optimizers. The
combined QMC--TNS framework therefore provides not only a simulator, but also
a generator of physically structured training data for intelligent
quantum-secure networks.

\section{Conclusions}

We have presented a broad quantum-inspired framework for adaptive
multi-demand routing optimization in QKD communication networks. The central
idea is to reformulate QKD routing as an interacting Hamiltonian system in
which latency, keyrate, risk, congestion and capacity limitations become
energetic contributions. Shared links generate effective interactions between
routing demands, producing a frustrated landscape analogous to sparse
interacting spin systems.

The framework combines two complementary simulation strategies. The
Quantum Monte Carlo inspired annealer performs stochastic thermal exploration
through local Metropolis moves and incremental Hamiltonian updates. The
stochastic Tensor-Network State method performs low-energy compression through
boundary-MPS or tensor-train truncation with thermal branch survival, random
demand ordering and a controllable bond dimension $\chi$. Together, these
methods diagnose not only the existence of low-energy routing configurations,
but also the compressibility, degeneracy and correlation complexity of the
routing landscape.

The implementation is deliberately formulated in terms of general
programming-environment variables and portable numerical structures: sparse
edge-index routes, precomputed path observables, persistent load vectors and
local energy updates. This avoids dependence on a specific language while
preserving the physical meaning of the model.

The resulting formulation connects QKD network routing with QUBO, Ising,
quantum annealing, QAOA, tensor-network optimization and future Quantum
Machine Learning systems. From a broader perspective, it provides a bridge
between quantum communication infrastructures, statistical physics,
stochastic simulation, variational compression and adaptive intelligent
network orchestration.

The proposed framework therefore establishes
a direct bridge between quantum communication networks,
statistical physics, tensor-network methods,
quantum-inspired optimization and future quantum-native
communication infrastructures
\cite{Kimble08,Wehner18,Preskill18}.

\section*{Acknowledgments}

This work has been supported by the ``Hub Nacional de Excelencia en
Comunicaciones Cuánticas'' project, funded by the Ministerio para la
Transformación Digital y de la Función Pública within the framework of the
Plan de Recuperación, Transformación y Resiliencia and the Mecanismo de
Recuperación y Resiliencia. Funded by the European Union --
NextGenerationEU.

\pagebreak
\appendix
\section{Meta-code of the QKD Routing QMC--Metropolis Optimizer}
\begin{tcolorbox}[
	colback=gray!5,
	colframe=black,
	title={Meta-code: Stochastic Boundary Truncation},
	breakable
	]

\begin{algorithmic}[1]
	
	\State \textbf{Input:} number of nodes $n$, average connectivity $k$, number of demands $M$, number of candidate paths $q$, number of Monte Carlo steps $N_{\rm steps}$.
	\State \textbf{Input:} Hamiltonian weights $\alpha_{\rm lat}$, $\beta_{\rm rate}$, $\gamma_{\rm risk}$, $\lambda_{\rm cong}$, $\mu_{\rm cap}$.
	\State \textbf{Input:} annealing parameters $\beta_0$, $\beta_F$ and saving interval $N_{\rm save}$.
	
	\vspace{0.2cm}
	\State \textbf{Construct a connected QKD graph}
	\State Set $m \simeq nk/2$.
	\Repeat
	\State Generate a random undirected graph $G=(V,E)$ with $n$ vertices and $m$ edges.
	\Until{$G$ is connected}
	\State Store the edge list $E=\{e_1,\ldots,e_{|E|}\}$.
	\State Build an edge-index map $e \mapsto i(e)$ for fast sparse lookup.
	
	\vspace{0.2cm}
	\State \textbf{Assign physical and operational link parameters}
	\For{each edge $e$}
	\State Sample an optical length $\ell_e$.
	\State Compute latency $\tau_e = c_\tau \ell_e$.
	\State Compute secret key generation rate $r_e = r_0 \exp(-\ell_e/\ell_0)+\eta_e$.
	\State Assign capacity $c_e = r_e+\xi_e$.
	\State Assign operational/security risk $\rho_e$.
	\EndFor
	
	\vspace{0.2cm}
	\State \textbf{Generate multi-demand traffic}
	\For{$a=1,\ldots,M$}
	\State Sample source node $s_a$ and target node $t_a$.
	\State Sample demand flow $f_a$.
	\EndFor
	
	\vspace{0.2cm}
	\State \textbf{Generate candidate routes}
	\For{$a=1,\ldots,M$}
	\State Compute up to $q$ candidate paths from $s_a$ to $t_a$.
	\If{fewer than $q$ paths are found}
	\State Complete the list by repeating the last available path.
	\EndIf
	\For{$p=1,\ldots,q$}
	\State Convert vertex path $P_{a,p}$ into sparse edge-index list
	\[
	\mathcal E_{a,p}=\{i(e): e\in P_{a,p}\}.
	\]
	\EndFor
	\EndFor
	
	\vspace{0.2cm}
	\State \textbf{Precompute route observables}
	\For{$a=1,\ldots,M$}
	\For{$p=1,\ldots,q$}
	\State Compute route latency:
	\[
	\tau_{a,p}=\sum_{e\in \mathcal E_{a,p}}\tau_e.
	\]
	\State Compute effective route keyrate:
	\[
	r_{a,p}=\min_{e\in \mathcal E_{a,p}} r_e.
	\]
	\State Compute effective route capacity:
	\[
	c_{a,p}=\min_{e\in \mathcal E_{a,p}} c_e.
	\]
	\State Compute route risk:
	\[
	\rho_{a,p}=\sum_{e\in \mathcal E_{a,p}}\rho_e.
	\]
	\EndFor
	\EndFor
	
	\vspace{0.2cm}
	\State Normalize $\tau_{a,p}$, $r_{a,p}$ and $\rho_{a,p}$ over all demands and candidate paths.
	
	\vspace{0.2cm}
	\State \textbf{Define the local routing energy}
	\For{$a=1,\ldots,M$}
	\For{$p=1,\ldots,q$}
	\State Set
	\[
	h_{a,p}
	=
	h^{\rm lat}_{a,p}
	+
	h^{\rm rate}_{a,p}
	+
	h^{\rm risk}_{a,p}
	+
	h^{\rm cap}_{a,p}.
	\]
	\EndFor
	\EndFor
	
	\vspace{0.2cm}
	\State \textbf{Initialize QMC state}
	\State Generate a random routing state
	\[
	\mathbf p=(p_1,\ldots,p_M), \qquad p_a\in\{1,\ldots,q\}.
	\]
	\State Compute the initial link-load vector
	\[
	L_e=\sum_{a=1}^{M} f_a\,\chi(e\in \mathcal E_{a,p_a}).
	\]
\State Compute the initial Hamiltonian.
\Statex
\[
\begin{aligned}
	H(\mathbf p)
	&=
	\sum_{a=1}^{M} h_{a,p_a}
	+
	\lambda_{\rm cong}\sum_e L_e^2
	\\
	&\quad
	+
	\mu_{\rm cap}
	\sum_e
	\max(0,L_e-c_e)^2 .
\end{aligned}
\]

\State Store the best initial configuration:
\Statex
\[
\mathbf p_{\rm best}=\mathbf p,
\qquad
L_{\rm best}=L,
\qquad
H_{\rm best}=H .
\]

\vspace{0.2cm}
\State \textbf{Metropolis--QMC annealing loop}

\For{$t=1,\ldots,N_{\rm steps}$}
\State Update inverse temperature.
\Statex
\[
\begin{aligned}
	\beta(t)
	&=
	\beta_0
	\left(
	\frac{\beta_F}{\beta_0}
	\right)^{
		\frac{t-1}{N_{\rm steps}-1}
	}.
\end{aligned}
\]
	\State Select a random demand $a$.
	\State Let $p_{\rm old}=p_a$.
	\State Select a new route $p_{\rm new}\neq p_{\rm old}$.
	\State Retrieve old and new edge-index lists:
	\[
	\mathcal E_{\rm old}=\mathcal E_{a,p_{\rm old}},
	\qquad
	\mathcal E_{\rm new}=\mathcal E_{a,p_{\rm new}}.
	\]
	\State Compute the local energy variation:
	\[
	\Delta H_{\rm local}
	=
	h_{a,p_{\rm new}}-h_{a,p_{\rm old}}.
	\]
	\State Identify the affected edge set:
	\[
	\mathcal A=\mathcal E_{\rm old}\cup \mathcal E_{\rm new}.
	\]
	\State For each affected edge, compute the load change induced by removing $f_a$ from the old route and adding $f_a$ to the new route.
	\State Compute the congestion/capacity variation:
	\[
	\Delta H_{\rm load}
	=
	\sum_{e\in\mathcal A}
	\left[
	\Phi_e(L_e+\Delta L_e)-\Phi_e(L_e)
	\right],
	\]
	where
	\[
	\Phi_e(x)
	=
	\lambda_{\rm cong}x^2
	+
	\mu_{\rm cap}\max(0,x-c_e)^2.
	\]
	\State Set
	\[
	\Delta H=\Delta H_{\rm local}+\Delta H_{\rm load}.
	\]
	\If{$\Delta H\leq 0$}
	\State Accept the move.
	\ElsIf{$u<\exp[-\beta(t)\Delta H]$, with $u\sim U(0,1)$}
	\State Accept the move.
	\Else
	\State Reject the move.
	\EndIf
	\If{move is accepted}
	\State Set $p_a=p_{\rm new}$.
	\State Update only the affected components of the load vector.
	\State Set $H\leftarrow H+\Delta H$.
	\EndIf
	\If{$H<H_{\rm best}$}
	\State Store $\mathbf p_{\rm best}=\mathbf p$.
	\State Store $L_{\rm best}=L$.
	\State Store $H_{\rm best}=H$.
	\EndIf
	\If{$t$ is a saving step}
	\State Store $(t,\beta(t),H,H_{\rm best})$ in the energy history.
	\EndIf
	\EndFor
	
	\vspace{0.2cm}
	\State \textbf{Output:} best energy $H_{\rm best}$, best routing state $\mathbf p_{\rm best}$, final load vector $L_{\rm best}$ and energy history.
	
\end{algorithmic}

\end{tcolorbox}
\pagebreak
\section{Meta-code of the Adaptive Minimum-Congestion Routing Module}

\begin{tcolorbox}[
	colback=gray!5,
	colframe=black,
	title={Meta-code: Stochastic Boundary Truncation},
	breakable
	]

\begin{algorithmic}[1]
	
	\State \textbf{Input:} optimized load vector $L_{\rm best}$ obtained from the QMC annealer.
	\State \textbf{Input:} source node $A$, destination node $B$ and new QKD message flow $f_{\rm new}$.
	\State \textbf{Input:} marginal congestion weights $\lambda_{\rm marg}$, $\mu_{\rm marg}$ and numerical offset $\epsilon$.
	
	\vspace{0.2cm}
	\State \textbf{Define marginal congestion energy}
	\For{each edge $e$}
	\State Define
	\[
	\Psi_e(x)
	=
	\lambda_{\rm marg}x^2
	+
	\mu_{\rm marg}\max(0,x-c_e)^2.
	\]
	\State Compute the marginal cost of sending the new QKD message through edge $e$:
	\[
	w_e
	=
	\Psi_e(L^{\rm best}_e+f_{\rm new})
	-
	\Psi_e(L^{\rm best}_e)
	+
	\epsilon.
	\]
	\EndFor
	
	\vspace{0.2cm}
	\State \textbf{Construct congestion-weighted graph}
	\State Build a weighted graph $G_{\rm cong}$ with the same vertices and edges as the original graph, but with edge weights $w_e$.
	
	\vspace{0.2cm}
	\State \textbf{Find the minimum-congestion route}
	\State Compute
	\[
	P_{\rm cong}(A,B)
	=
	\arg\min_{P:A\rightarrow B}
	\sum_{e\in P} w_e.
	\]
	\State Convert the path into its edge-index representation.
	\State Compute the total marginal congestion cost:
	\[
	C_{\rm cong}
	=
	\sum_{e\in P_{\rm cong}} w_e.
	\]
	
	\vspace{0.2cm}
	\State \textbf{Update hypothetical load}
	\State Define a new load vector
	\[
	L'_e =
	\begin{cases}
		L^{\rm best}_e+f_{\rm new}, & e\in P_{\rm cong},\\
		L^{\rm best}_e, & e\notin P_{\rm cong}.
	\end{cases}
	\]
	
	\vspace{0.2cm}
	\State \textbf{Compare with the topological shortest path}
	\State Compute the unweighted shortest path $P_{\rm topo}(A,B)$ in the original graph.
	\State Compute its marginal congestion cost:
	\[
	C_{\rm topo}
	=
	\sum_{e\in P_{\rm topo}} w_e.
	\]
	\State Compute the relative congestion reduction:
	\[
	R
	=
	100\left(
	1-\frac{C_{\rm cong}}{C_{\rm topo}}
	\right)\%.
	\]
	
	\vspace{0.2cm}
	\State \textbf{Output:} minimum-congestion path $P_{\rm cong}$, topological shortest path $P_{\rm topo}$, costs $C_{\rm cong}$ and $C_{\rm topo}$, updated load vector $L'$ and relative congestion reduction $R$.
	
\end{algorithmic}

\end{tcolorbox}
\section{Meta-code of the Stochastic Tensor Truncation Rule}
\begin{tcolorbox}[
	colback=gray!5,
	colframe=black,
	title={Meta-code: Stochastic Boundary Truncation},
	breakable
	]
	
	\begin{algorithmic}[1]
		
		\Require Expanded tensor boundary $\mathcal B=\{s_i\}_{i=1}^{N_B}$
		\Require Bond dimension $\chi$
		\Require Tensor inverse temperature $\beta_{\rm TNS}$
		\Require Noise amplitude $\epsilon_{\rm TNS}$
		
		\For{each branch $s_i\in\mathcal B$}
		\State Read branch energy $E_i$.
		\State Add a small stochastic perturbation.
		\Statex
		\[
		\tilde E_i
		=
		E_i+\eta_i,
		\qquad
		\eta_i\sim U(-\epsilon_{\rm TNS},\epsilon_{\rm TNS}) .
		\]
		\EndFor
		
		\State Compute the minimum perturbed energy.
		\Statex
		\[
		\tilde E_{\min}=\min_i \tilde E_i .
		\]
		
		\For{each branch $s_i\in\mathcal B$}
		\State Assign Boltzmann-like survival weight.
		\Statex
		\[
		w_i=
		\exp\!\left[
		-\beta_{\rm TNS}
		\left(
		\tilde E_i-\tilde E_{\min}
		\right)
		\right] .
		\]
		\EndFor
		
		\State Normalize the weights.
		\Statex
		\[
		P_i=\frac{w_i}{\sum_j w_j}.
		\]
		
		\State Randomly sample up to $3\chi$ candidate branches according to probabilities $P_i$.
		\State Remove duplicate branches.
		\State Sort the selected branches by their unperturbed physical energy $E_i$.
		\State Retain the lowest-energy $\chi$ branches.
		
		\Ensure Truncated tensor boundary $\mathcal S$ with at most $\chi$ retained branches.
		
	\end{algorithmic}
	
\end{tcolorbox}
\noindent
\pagebreak
\section{Meta-code of the TNS Adaptive Minimum-Congestion Rerouting Module}
\begin{tcolorbox}[
	colback=gray!5,
	colframe=black,
	title={Meta-code: Adaptive Minimum-Congestion Routing after TNS Optimization},
	breakable
	]
	
	\begin{algorithmic}[1]
		
		\Require Final TNS load vector $L^{\rm TNS}$
		\Require Source node $A$, destination node $B$ and new QKD message flow $f_{\rm new}$
		\Require Marginal congestion weights $\lambda_{\rm marg}$, $\mu_{\rm marg}$ and offset $\epsilon$
		
		\For{each edge $e$}
		\State Define the marginal congestion energy.
		\Statex
		\[
		\Psi_e(x)
		=
		\lambda_{\rm marg}x^2
		+
		\mu_{\rm marg}\max(0,x-c_e)^2 .
		\]
		
		\State Compute the marginal cost of routing the new message through edge $e$.
		\Statex
		\[
		w_e
		=
		\Psi_e\!\left(L^{\rm TNS}_e+f_{\rm new}\right)
		-
		\Psi_e\!\left(L^{\rm TNS}_e\right)
		+
		\epsilon .
		\]
		\EndFor
		
		\State Construct a weighted graph $G_{\rm cong}$ with the same topology as the original QKD graph and with edge weights $w_e$.
		
		\State Compute the minimum-congestion path.
		\Statex
		\[
		P_{\rm cong}(A,B)
		=
		\arg\min_{P:A\rightarrow B}
		\sum_{e\in P} w_e .
		\]
		
		\State Compute the corresponding marginal congestion cost.
		\Statex
		\[
		C_{\rm cong}
		=
		\sum_{e\in P_{\rm cong}} w_e .
		\]
		
		\State Compute the ordinary topological shortest path $P_{\rm topo}(A,B)$ in the unweighted graph.
		
		\State Compute its associated congestion cost.
		\Statex
		\[
		C_{\rm topo}
		=
		\sum_{e\in P_{\rm topo}} w_e .
		\]
		
		\State Compute the relative congestion reduction.
		\Statex
		\[
		R
		=
		100
		\left(
		1-\frac{C_{\rm cong}}{C_{\rm topo}}
		\right)\% .
		\]
		
		\Ensure Minimum-congestion path $P_{\rm cong}$, topological path $P_{\rm topo}$, marginal costs $C_{\rm cong}$ and $C_{\rm topo}$, and relative congestion reduction $R$.
		
	\end{algorithmic}
	
\end{tcolorbox}

\end{document}